\documentclass[11pt, epsfig]{article}
\usepackage{epsfig, amsmath, amssymb, amsthm, times}
\theoremstyle{defintion}

\theoremstyle{remark}

\def\lbl{\label}
\def\be{\begin{equation}}
\def\ee{\end{equation}}
\def\lbl{\label}

\def\E{{\mathbb E}}

\def\Re{{\mathbb R}}

\title{Chaos at the border of criticality}
\author{ Georgi S. Medvedev and Yun Yoo
\thanks{
Department of Mathematics, Drexel University, 3141 Chestnut Street,
Philadelphia, PA 19104, {\tt medvedev@drexel.edu} 
}
}
\begin{document}
\maketitle
\begin{abstract}
The present paper points out to a novel scenario for formation of chaotic
attractors in a class of models of excitable cell membranes near an Andronov-Hopf 
bifurcation (AHB). The mechanism underlying chaotic dynamics admits a simple and visual description
in terms of the families of one-dimensional first-return maps, which 
are constructed using the combination of asymptotic and numerical techniques. 
The bifurcation structure of the continuous system (specifically, the proximity
to a degenerate AHB) endows the Poincare map with distinct
qualitative features such as unimodality and the presence of the boundary layer,
where the map is strongly expanding. This structure of the map in turn explains the 
bifurcation scenarios in the continuous system including 
chaotic mixed-mode oscillations near the border between the regions of sub-
and supercritical AHB. The proposed mechanism yields the statistical properties of 
the mixed-mode oscillations in this regime. The statistics predicted by the analysis
of the Poincare map  and those observed in the numerical experiments of the continuous
system show a very good agreement.
\end{abstract}

\vfill
\newpage
{ \bf Identifying bifurcation scenarios leading to the formation of chaotic attractors
is one of the central problems of nonlinear science. In applications, such information can
be used for locating and characterizing the regions of complex behavior. The present paper
identifies the proximity to a degenerate Andronov-Hopf bifurcation as a source of 
complex dynamics in differential equation models of Hodgkin-Huxley type. The latter constitute
an important class of models of computational biology. 
The results of this work explain the origin of chaotic mixed-mode oscillations generated
by models of this class and yield the statistical properties of the irregular oscillatory patterns.  
The statistics predicted by the analysis show a very good agreement with those
estimated in the numerical experiments.
}
\vfill
\newpage

\setcounter{equation}{0}
\section{Introduction}
Understanding mechanisms responsible for generating different patterns of electrical
activity in neurons, firing patterns, and transitions between them is fundamental for
understanding how the nervous system processes information. After a classical series of
papers by Alan Hodgkin and Andrew Huxley \cite{HH}, nonlinear differential equations became
a common framework for modeling electrical activity in neural cells. Today the language
and techniques of the dynamical systems theory, especially those of the bifurcation theory, 
are an indispensable part of understanding computational biology.
A distinctive feature of many biological models is that they are often close to
a bifurcation. In particular, all excitable systems, including many models of neural cells
reside near a bifurcation. The type of the bifurcation involved in the mechanism of 
action potential generation underlies a fundamental classification of neuronal 
excitability \cite{RE}.
According to this classification, the excitation in type II models is realized 
via an AHB. The proximity of  a model to an AHB can have 
a significant impact on the firing patterns that they produce. Near the bifurcation, systems 
acquire greater flexibility in generating dynamical patterns varying in form and frequency. 
The proximity to the instability also makes this class of systems more likely to exhibit chaotic behavior. 
Typical mechanisms of transitions to chaotic dynamics in conductance-based models of neural 
cells are much less understood compared to those corresponding to periodic activity. 
Understanding these mechanisms is important in view of the prevalence of irregular dynamics 
in both modeling and experimental studies of neural cells.

\begin{figure}
\begin{center}
\epsfig{figure=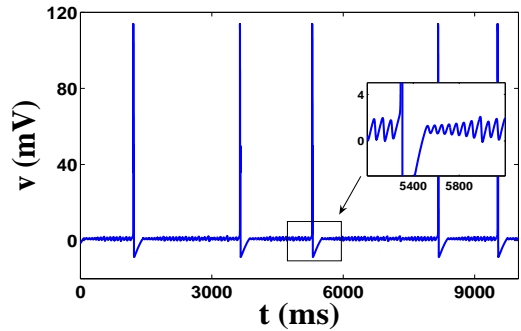, height=1.8in, width=2.8in, angle=0}
\end{center}
\caption{
The time series of $v(t)$ of the modified HH model (for the parameter values, see Appendix).
In the present regime, the
equilibrium of (\ref{1}), (\ref{2}) is near the AHB. The bifurcation is 
subcritical but is very close to the border, where the type of the bifurcation
changes to a supercritical. The proximity of the AHB to the border of criticality
combined with the slow-fast structure of (\ref{1}), (\ref{2}) results in the
irregular MMOs shown in this figure. 
}\label{f.1}
\end{figure}

In the present paper, we describe a regime of chaotic mixed-mode oscillations (MMOs)
in type II conductance based models. Mixed-mode oscillatory patterns combine small
amplitude (subthreshold) oscillations with large amplitude spikes (Fig. \ref{f.1}).
Despite subtle underlying dynamical structures,  
MMOs have been reported in many experimental and modeling studies of neural cells 
\cite{DK, DNK, DRSE, MC, RW, RW_chaos} as well as in the models of phenomena outside of neuroscience
\cite{FKRT, MSLG, VV}. The mathematical mechanisms generating MMOs have recently become an
area of intense research. Several structures have been proposed to serve as organizing centers
for MMOs. Among them, there are folded singularities \cite{BKW, MS, WESCH}, proximity to the 
AHB combined with the strongly contracting return mechanism \cite{MY}, and the Hopf-homoclinic
bifurcation \cite{GW}. Recent studies have greatly elucidated the sources for periodic mixed-mode
regimes. The mechanisms for irregular MMOs remain to be explained.  The present paper suggests
a general mechanism for chaotic MMOs in terms of the bifurcation structure of 
the corresponding class of models. Specifically, we show that the transition from supercritical to
subcritical AHB in a class of systems lies through the regime of chaotic MMOs. Consequently,
systems lying close to such transition are likely to exhibit chaotic behavior. This effect
was first noticed in the context of the model of solid fuel combustion in (both in the original infinite
dimensional model and its finite dimensional approximation) \cite{FKRT}. 
The qualitative explanation for the mechanism of chaos near the border of criticality (i.e., where
the AHB changes its type from super- to subcritical)
for a class of systems, which includes the model in \cite{FKRT} was given in \cite{MY}. In the 
present paper, we show that the  necessary ingredients for this scenario are naturally embedded 
in the structure of type II conductance based models of excitable cells. The latter covers
many important biological models, including the 
Hodgkin-Huxley (HH) model (which we use below to illustrate the proposed mechanism) among many other 
conductance based models of neurons.
We propose a simple geometric mechanism 
that explains the origins of chaotic dynamics, using a family of one-dimensional ($1D$)
unimodal first return maps (Fig. \ref{f.3}).
The qualitative features of the Poincare maps such as unimodality and the presence
of the boundary layer, where the map is strongly expanding, follow from the proximity
of the continuous system to a degenerate AHB and the slow-fast character of the global
vector field. Therefore, the bifurcation scenario and the mechanism of chaotic oscillations
described below, are independent from the specific details of the model, but are determined
by the bifurcation structure pertinent to a whole class of models. 
Furthermore, we complement and confirm the conclusions of \cite{MY} and of this paper
on the presence of the chaotic attractors in the corresponding families of models 
by computing the statistics associated with the irregular MMOs. 
The latter clearly show 
that the timings of spikes in the mixed mode patterns are distributed geometrically
in accord with our theoretical estimates.

\section{ The model}
A large class of models of excitable cell membranes relies on the conductance based
formalism due to Hodgkin and Huxley \cite{HH}. Despite a rich diversity
featured by different mathematical models used to elucidate the electrophysiology of various 
neural cells, many of them share the same structure reminiscent of the classical
HH model. Specifically, a typical differential equation model of a (point) neuron consists
of a current balance equation and the equations for gating variables, which often
present disparate time scales. In identifying new dynamical structures generated by the 
models of this class, the specific choices of the parameter values in a model at hand 
may not be so important (unless one is interested in a particular cell type), as they change 
from one model to another and also depend on many other factors (such as temperature, exposure
to neuroactive substances, etc). However, it is critical to preserve the
overall structure common to conductance-based models. This is the motivation and 
the justification for the choice of the model for the present study. The model to be introduced
below should not be viewed as a model of a certain neuron, but rather as
a representative example of a conductance based model.
We consider a modified HH model introduced by Doi and Kumagai \cite{DK}.
This is a system of nonlinear ordinary differential 
equations for the membrane potential, $v$, and two gating variables 
$h$ and $n$:
\begin{eqnarray} \lbl{1}
C_m\dot v &=& F(v,h,n,I_{ap}),\\ 
\lbl{2}
\bar \tau_s \dot s &=& \frac{s_\infty(v)-s}{\tau_s(v)}, \; s \in \{h, n\},
\end{eqnarray}
where $F(v,h,n,I_{ap}):=-g_{Na}m_{\infty}(v)^3 h(v-E_{Na})-g_{K}n^4(v-E_{K})-g_{L}(v-E_{L})+I_{ap}$ 
is the sum of ionic currents and applied current $I_{ap}$.
Constants $E_c$ and $g_c,$ $c\in\left\{Na, K, L\right\}$ stand for reversal potentials
and maximal conductances of the corresponding ionic currents.
Functions $m_\infty(v), s_\infty(v)$ and $\tau_s(v),$ $s\in\left\{ h,n\right\}$ are used
in the descriptions of the ionic channels' kinetics and have typical for the HH model
forms.
For the analytic expressions of these functions and the values of parameters, 
we refer the reader to the Appendix to this paper (see also \cite{DK}).
The only approximation used in the original HH model to obtain (\ref{1}),(\ref{2}) 
is the steady state approximation for gating variable $m \approx m_{\infty}(v)$, 
which uses the separation of the timescales in the HH model to eliminate the equation 
for $m$. A mathematical justification of this approximation via a center manifold
reduction can be found in \cite{RW}. 
Following \cite{DK}, we view  $\bar \tau_{h,n}$
as control parameters, thus, allowing for a range of timescales in (\ref{1}) and (\ref{2}).
Note that ionic conductances generating electrical 
activity in different cells have widely varying time constants. 
Therefore, treating $\bar\tau_{h,n}$ as control parameters 
allows one to investigate the qualitative dynamics for a 
class of conductance based models, while preserving the 
biophysical structure of the HH model.
In particular, the modified HH model proved very useful for studying mechanisms for MMOs
generated by conductance based models \cite{DK, RW, RW_chaos}. 
The numerical simulations of the modified HH model \cite{DK, DNK}
revealed multiple parameter configurations producing chaotic MMOs. 
Below, we describe a general mechanism responsible for
generating chaos in this and similar models.

\section{The Poincare map}
We start by describing the qualitative structure of the model.
From (\ref{1}) and (\ref{2}) one easily finds the equations for the fixed points:
\begin{equation} \lbl{3}
F(\bar v, h_\infty(\bar v),n_\infty(\bar v),I_{ap})=0, \; \bar s=s_\infty(\bar v),\; s\in\{h, n\}.
\end{equation}
In the range of parameters of interest, system of equations (\ref{1}),(\ref{2}) 
has a unique fixed point residing near 
the AHB. In the present study, we fix $\bar \tau_h=1$ and choose 
$I_{ap}$ 
so that the linearization of 
(\ref{1}),(\ref{2}) 
around $(\bar v, \bar h, \bar n)$ has three eigenvalues:
\begin{equation}\label{4}
\lambda < 0 \;\;\mbox{and}\;\; \alpha \pm \beta i,
\end{equation}
where $\quad 0 < \alpha \ll 1, \beta>0.$
Thus, (\ref{1}),(\ref{2}) is close to the AHB. We will use $\bar\tau_n$ to control the type of the
AHB (sub- or supercritical). By a linear change of variables, (\ref{1}),(\ref{2}) can be rewritten as
\be\lbl{5}
\left(\begin{array}{c}
\dot \xi\\ \dot \eta_1\\ \dot \eta_2
      \end{array}\right) =
\left(\begin{array}{ccc}
                      \lambda & 0 & 0\\ 
                       0 & \alpha & -\beta\\ 
                       0 & \beta &  \alpha
      \end{array}\right)
\left(
       \begin{array}{c}
                \xi\\  \eta_1\\ \eta_2
      \end{array}\right)      +
{\cal N}(\xi,\eta),
\ee
where ${\cal N}$ and $D{\cal N}$ vanish at the origin.
\begin{figure}
\begin{center}
\epsfig{figure=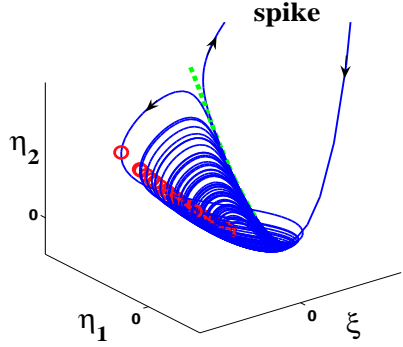, height=2.0in, width=2.4in, angle=0}
\end{center}
\caption{
A part of the phase trajectory of (\ref{5}) is plotted during 
the phase of the small oscillations between two successive spikes.
Prior to leaving a neighborhood of the origin the trajectory
remains close to a $2D$ slow manifold. 
The intersection of the 
slow manifold with cross-section $\Sigma$ (used for the
construction of the Poincare map) is indicated by the
line of circles. For computational convenience, in the 
present study we assumed that $\Sigma$ intersects the
slow manifold along the curve of maxima of $\eta_1$.
Other suitable choices of $\Sigma$ are expected to yield
qualitatively similar results.
}\label{f.2}
\end{figure}
Near the origin, (\ref{5}) has a $1D$ local stable manifold $W_{loc}^s(O)$ and $2D$ 
local weakly unstable manifold, $W_{loc}^u(O)$.

To describe the mechanism for chaotic oscillations in (\ref{5}), we refer
to the plot of a typical phase trajectory near the origin. 
The trajectory shown in Fig. \ref{f.2} approaches a weakly unstable equilibrium at the 
origin very closely along $W^s(O)$ and then spirals out along $W^u(O)$. Just near the origin
the oscillatory dynamics can be understood via the local analysis of (\ref{5}) (cf. \cite{MY}).
The oscillations of the intermediate amplitude shown in Fig. \ref{f.2} display distinct
canard structure. Note that different rings of the spiraling trajectory lie very closely to 
each other along the dashed curve. 
The dashed curve in Fig. \ref{f.2} indicates the canard trajectory separating small and intermediate oscillations
from large amplitude spikes \cite{MS}. Once the amplitude of oscillations gets sufficiently large, 
at the end of the long excursion along the dashed curve, the
trajectory either escapes from a small neighborhood of the origin and generates a spike,
or makes another cycle of oscillations near the origin. The mathematical analysis of canard
trajectories in $\Re^3$ and MMOs in general is a difficult and technical problem 
(see \cite{BKW, MY, MS, RW, RW_chaos, WESCH}). 
The irregular MMOs such as shown in Fig. \ref{f.2} have not been analyzed yet.
We will address the mathematical analysis of this problem in the future work. 
In the present paper, we rely
on the numerical observations outlined above, which suggest that near the origin the
trajectories remain close to a $2D$ slow manifold. 
Therefore, the corresponding dynamics can be 
analyzed with a $1D$ Poincare map. Below, we construct such
map using the combination of the analytical and numerical techniques.    
\begin{figure}
\begin{center}
{\bf a}\;\epsfig{figure=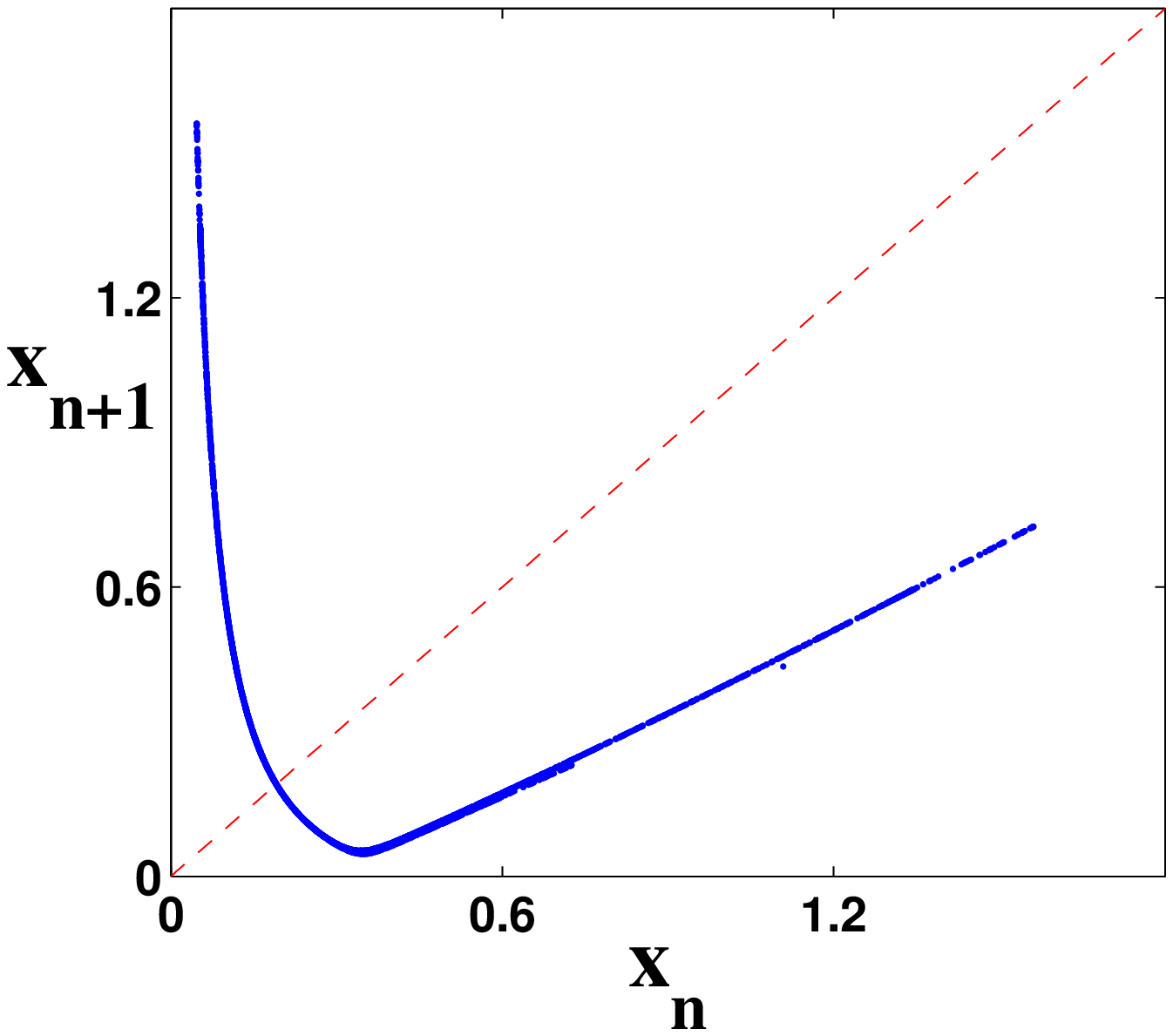, height=1.8in, width=2.0in, angle=0}
{\bf b}\;\epsfig{figure=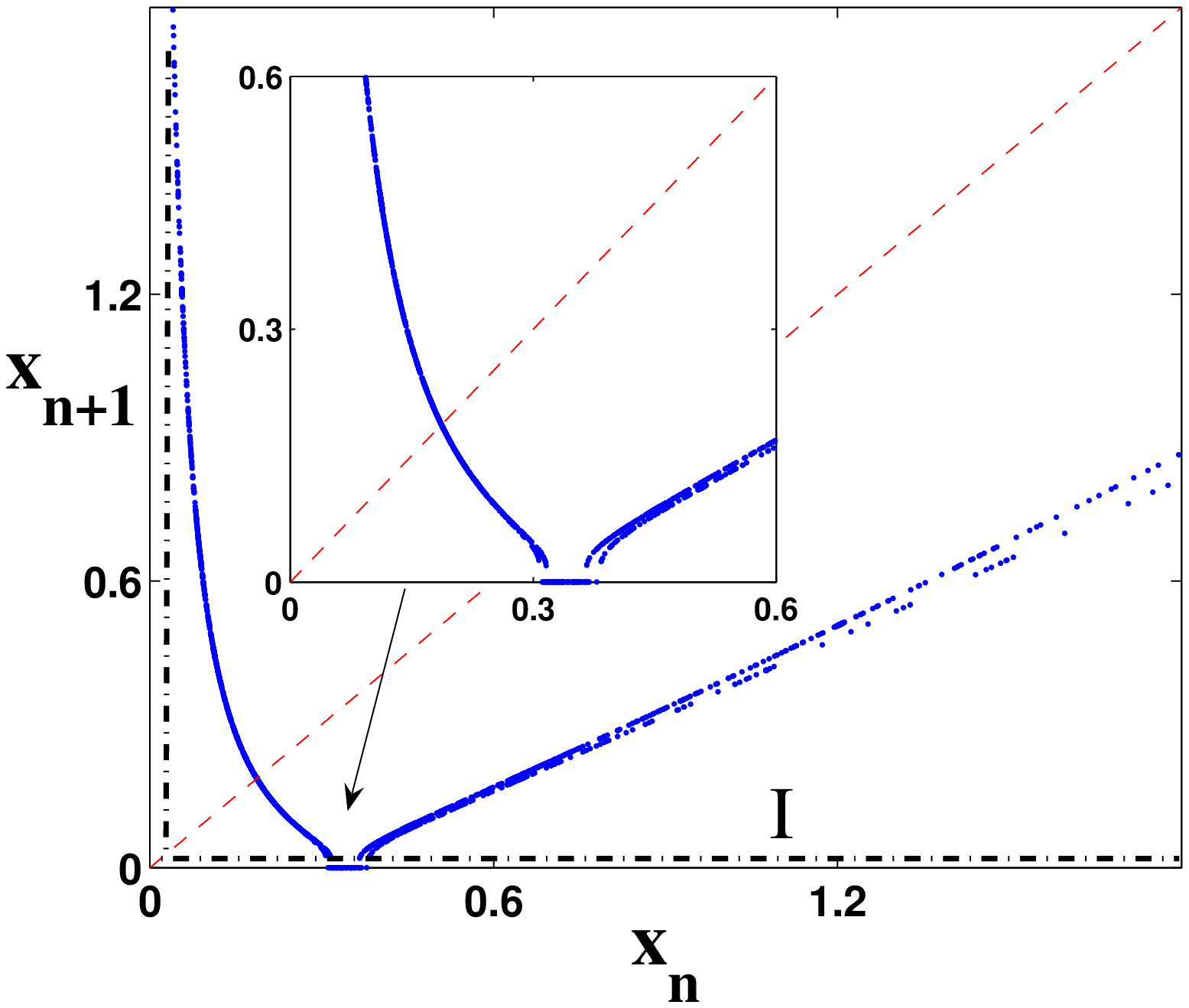, height=1.8in, width=2.0in, angle=0}
\end{center}
\caption{The Poincare maps plotted for two values of the control parameter
$\bar\tau_n:$ $43.02$ (a) and $42.88$ (b). The values of $\bar\tau_n$ are chosen so that 
the system is close to the border between the regions of sub- and supercritical 
AHB.
The values of the other parameters are the same as in the Appendix.
}\label{f.3}
\end{figure}

To this end, we introduce a cross-section, $\Sigma$, transverse to the slow manifold, and
look at the successive intersections of the trajectories of (\ref{5}) with $\Sigma$
(Fig. \ref{f.2}).
The Poincare map has especially convenient form in terms of 
\be\lbl{x}
x=\alpha(\eta_1^2+\eta_2^2)^{-1}.
\ee
Specifically, we define
\be\lbl{6}
P:x(t_i) \mapsto x(t_{i+1}), \; i=1,2,...,
\ee
where $t_i$ are the times of successive crossings of the trajectory of (\ref{1}), (\ref{2})
with $\Sigma$. The numerically computed maps for two values of $\bar\tau_n$ are shown in Fig. \ref{f.3}. 
Note that the small
values of $x$ correspond to the large amplitude oscillations of (\ref{5}). 
We are interested in the oscillations of small and intermediate amplitudes. Therefore, 
in the domain of $P$ we distinguish the  region bounded away from some $O(\alpha)$ neighborhood
of the origin, $I$, which supports small amplitude oscillations. We call $I$ the principal domain 
of $P$ (see Fig. \ref{f.3}b). In $I$, $P$ has a distinct layered structure. In the outer region, bounded
away from the origin, the map is almost linear. 
In the complementary boundary layer, the map is unimodal 
with the slope becoming increasingly steep near the origin (Fig. \ref{f.3}a,b). 
The almost linear form of the map in the outer region follows from the local analysis of (\ref{5})
near the origin (cf. \cite{MY}):
\be\lbl{7}
P(x)=x-2\alpha\omega(x+\gamma)+O(\alpha^2), \; \omega=2\pi\beta^{-1},
\ee
where $\alpha$ and $\beta$ are given in (\ref{4}).
Note that remarkably simple form of the map in (\ref{7}) is afforded by the choice of the 
independent variable $x$ (\ref{x}).
The first Lyapunov coefficient $\gamma$ determines the type of the AHB. The bifurcation
is subcritical (supercritical) if $\gamma >0 (\gamma<0)$.
The dynamics of $P$ in the boundary layer corresponds to the canard like oscillations 
of intermediate amplitude in (\ref{5}). The formal asymptotic expression for
$P$ in the boundary layer can be found in \cite{MY}. 
Providing a rigorous analytical description of $P$ in the boundary layer is beyond the scope
of this paper: it requires the information about the canard cycles of (\ref{5}). 
For the purposes of the present study, 
we refer to the numerical results in Fig. \ref{f.3}, which provide a clear evidence 
in favor of the unimodality 
of $P$ in the boundary layer. The latter combined with the additive dependence of $P$ on $\gamma$
(see (\ref{7})) suggests the mechanism for chaotic MMOs in (\ref{5}) in the regime
$\gamma=O(\alpha)$, i.e. near the border of criticality of the AHB.
\begin{figure}
\begin{center}
\epsfig{figure=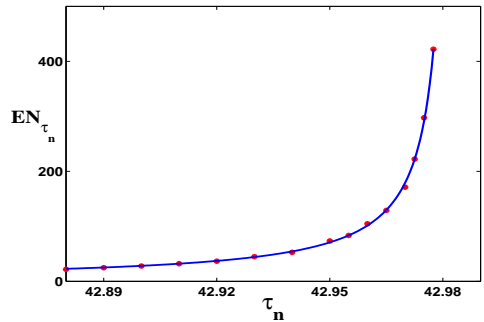, height=1.8in, width=2.8in, angle=0}
\end{center}
\caption{
The expected value of the number of small amplitude oscillations, $\E N_{\bar\tau_n}$ as a function 
of $\tau_n$. Numerical data
are fitted according to theoretical estimate (\ref{8}) with
$a_2\approx 2.34$  and $\bar\tau^1_n\approx 42.983.$ 
}\label{f.4}
\end{figure}

Note that for negative $\gamma=O(1)$, (\ref{6}) has a unique fixed point, which lies in the
outer region (see (\ref{7}) for the description of $P$ in the outer region).
This fixed point corresponds to
a stable periodic orbit of (\ref{5}) born at a supercritical AHB. For increasing values of 
$\gamma$, by (\ref{6}),
the graph of the map moves down and the fixed point looses stability via a period-doubling bifurcation
as it enters the boundary layer (Fig. \ref{f.3}a).
This triggers the PD cascade (not shown in this paper; see \cite{MY}, where it is shown for 
a closely related model) and leads to the formation of the chaotic attractor in accord with
the classical scenario known for unimodal maps \cite{GH}.
Now we are in a position to describe the mechanism for chaotic MMOs. Note that as long as 
$\min_{x\in I} P(x)$ remains bounded away from some  $O(\alpha)$ neighborhood of $0$ (see Fig. \ref{f.3}a),
$P(I)\subset I$ and the dynamics is trapped in $I$. This means that the amplitude of oscillations
can not get too big (recall the definition of $x$ in (\ref{x})). However, as $\gamma$ is approaching
$0$ ($\gamma=O(\alpha)$), the graph of $P(x)$ over $I$ moves down and at some point develops a small window
of escape (see the small horizontal segment in the graph of $P$ near
the former point of minimum of $P$,  Fig. \ref{f.3}b).
Now the irregular dynamics in $I$ can be interrupted if the trajectory hits the window of escape.
This event, corresponding to a spike in the continuous system (Fig. \ref{f.1}), is followed by the
reinjection of the trajectory back to $I$. This in turn initiates a new series of small oscillations.
Note that just near the transition from the subthreshold oscillations to MMOs, the size of the window 
is very small and a typical trajectory
stays in $I$ for a long time before escape. Thus, just near the transition point, the typical patterns
consist of large (random) number $N_{\bar\tau_n}$ of small amplitude oscillations between successive spikes.
Moreover, assuming that just before the transition, the dynamics in $I$ is mixing, one can show
that $N_{\bar\tau_n}$ is distributed geometrically and estimate the parameter of the geometric distribution 
(cf. \cite{M06}).
In the following section, we support the proposed scenario with results of numerical simulations.

\section{The statistics of irregular MMOs} 

In qualitative form, the mechanism generating irregular MMOs in (\ref{5}) via escape from the 
principal domain was studied in \cite{M06} in the context of chaotic bursting 
in a class of 
conductance based models. Note that the differential equation models considered
in \cite{M06} and in the present paper possess quite different bifurcation structures.
The affinity between the mechanisms for chaotic dynamics in these classes of models 
shows up only after the reduction to suitable $1D$ Poincare maps. 
While the basis for the reduction to the map is specific to each case, the qualitative
properties of the resultant maps are remarkably similar. In particular,
the analysis of the statistical
properties of the irregular bursting patterns in \cite{M06} carries over to that of chaotic MMOs. 
There are two main outcomes of the analysis in \cite{M06}. Translated to the
present problem, they can be formulated as follows. Suppose $\bar\tau_n^c$ is a critical value
of the control parameter, separating the subthreshold oscillations $(\bar\tau_n>\bar\tau_n^c)$ from
the MMOs $(\bar\tau_n<\bar\tau_n^c)$. 
Suppose, in addition that just before the transition to MMOs the dynamics is mixing. 
This assumption is motivated by the presence of the steep segment in the graph of the 
map in the boundary layer near $0$ (see Fig. \ref{f.3}), as well as by the results of direct simulations
of (\ref{1}),(\ref{2}).
Then for $\bar\tau_n^c-\bar\tau_n>0$ sufficiently small, the number
of subthreshold oscillation between successive spikes, $N_{\bar\tau_n}$, can be approximated
by a geometric random variable (cf. \cite{M06}). Moreover, by estimating the parameter of the geometric distribution,
one can predict the functional form of the dependence of the expected value $\E N_{\bar\tau_n}$ on 
the distance from the transition point $\bar\tau_n^c-\bar\tau_n>0$. In particular, 
the results of \cite{M06} imply that near the border of criticality, 
\begin{equation}\lbl{8}
\E N_{\bar\tau_n}\approx\left\{
\begin{array}{cc}
a_1\left(\bar\tau_n-\bar\tau^c_n\right)^{-1/2}, & \bar\tau_n^1\lessapprox\bar\tau_n<\bar\tau_n^c,\\
a_2\left(\bar\tau_n-\bar\tau^1_n\right)^{-1}, & \bar\tau_n<\bar\tau_n^1,
\end{array}
\right.
\ee
for some positive constants $a_1, a_2$ and  for a certain value of $\bar\tau_n^1$, which 
is less than critical value $\bar\tau_n^c$ but is sufficiently close to it. We verified these
prediction numerically for a set of values of $\bar\tau_n$ near the transition point.
In all cases, we found that $N_{\bar\tau_n}$ are distributed geometrically.
 A representative plot of the probability density
function is shown in Fig. \ref{f.5}. Moreover, the data in Fig. \ref{f.4} show a very good
fit with the theoretical estimate in the second line of (\ref{8}). We did not
attempt numerical verification of the estimate in the top line of (\ref{8}), because it requires
numerical integration of (\ref{1}),(\ref{2}) over extremely long intervals of time and is expected
to hold in a rather narrow interval of $\bar\tau_n$. 
In addition to verifying (\ref{8}), the numerical experiments presented in this section support
the claim that the dynamics of (\ref{1}),(\ref{2}) near the border of criticality is chaotic.
In particular, the fact that the distribution of $N_{\bar\tau_n}$ is geometric provides a posteriori
justification for the assumption that dynamics of $P$  is mixing in $I$, which was used in \cite{M06}
to infer that $N_{\bar\tau_n}$ is approximately a geometric random variable and to derive (\ref{8}). 
Therefore, the map-based description of the transformations in the dynamics of (\ref{1}) and 
(\ref{2}), which accompany the approach of the border between regions of supercritical and subcritical
AHB, combined with the numerical results presented in this section form a convincing evidence
of the existence of the chaotic attractor in this regime. We emphasize the universal character
of this effect: the principal ingredients necessary for its realization such as the proximity to a
degenerate AHB and the reinjection mechanism are expected to be present in a large class of systems.
In particular, the same mechanism is responsible for generating chaotic MMOs in a model 
of solid fuel combustion \cite{MY}. The results of the present study suggest that
it is relevant for an important class of biological models.
\begin{figure}
\begin{center}
\epsfig{figure=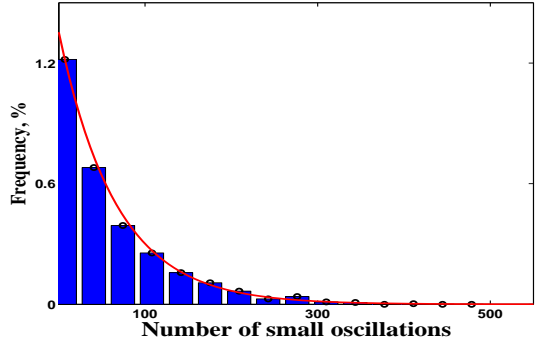, height=1.8in, width=2.8in, angle=0}
\end{center}
\caption{The probability distribution of the number of oscillations 
between two successive spikes in a typical trajectory of (\ref{1}),(\ref{2})
for $\bar\tau_n\approx 42.95.$ The histogram for $N_{\tau_n}$ is obtained using 
the numerical data for $10^3$ interspike intervals and is fitted
with the exponential probability density function approximating a shifted 
geometric distribution: $7+\mbox{Geom(66.5)}$ (plotted with solid line).  
}\label{f.5}
\end{figure}

\section{Discussion}

Substantial efforts have been made to identify and to understand the mechanisms
generating chaotic dynamics in physical and biological systems, and in particular,
in conductance based models of neural cells \cite{CR, GO, M06, RA, T}. 
In the context of neuronal modeling, 
understanding the sources of chaos is important for elucidating
the origins and characterizing the properties of patterns of irregular electrical
activity in the nervous system. The latter are quite common on both cellular
and network levels. There is a compelling evidence showing that irregular firing
in many cases results from the intrinsic properties of neural cells and
their networks, rather than from random fluctuations. Although, there is no complete
understanding for the functional role of chaos in neural systems, there are several lines
of evidence indicating its potential importance. First, there are modeling studies suggesting
that chaotic elements enhance the information processing properties of the
networks that they form \cite{RA}. Second, the loss of regularity in certain patterns of neural 
activity are known to accompany the onset of some pathological states, 
such as epileptic seizures and Parkinsonian states.
Therefore, better understanding of the origins of chaos in neuronal models may help to 
predict the onset of pathological states and to control them.
Single cell models such as (\ref{1}),(\ref{2}) provide the simplest biophysically meaningful
framework for studying chaos in neuronal dynamics. Our study supports the
view that chaotic dynamics is a likely attribute of a system operating on the edge of
instability \cite{RA}. Note that on the border of criticality, the regime of interest in
our study, type II models acquire remarkable pattern-forming capacity and 
flexibility: upon relatively small changes of parameters they exhibit
a rich repertoire of qualitatively different patterns ranging from subthreshold
oscillations (supercritical AHB) to a variety of periodic and aperiodic MMOs
(subcritical AHB) \cite{DK, MY}. Thus, the ability of the system to detect and to encode small
changes in the stimulus is greatly enhanced in this regime.
The present paper points out that in a broad class of type II models of 
neural cells, the proximity to the border of criticality of the underlying AHB, 
implies chaotic dynamics. Moreover, the results in \cite{MY} suggest that
under certain natural constraints, for small $\alpha >0$, this is the only
region in the parameter space of (\ref{1}), (\ref{2}), where robust chaotic oscillations
are expected. For this regime, we provide a geometric explanation for the mechanism of 
chaotic dynamics, using a family of $1D$ Poincare maps. We conducted a series of
numerical experiments to verify our theoretical predictions and found that
the former are in a good agreement with the latter. 
Our results can be used to locate chaotic regimes in type II models and to describe the 
statistical properties of the resultant oscillatory patterns. Finally, we emphasize that the results of
the present study apply to a class of conductance based models (for which (\ref{1}),(\ref{2})
is only a representative example), as well as to models outside biology. In particular, we have
obtained qualitatively similar results 
(including the statistical data as shown in Fig. \ref{f.4} and \ref{f.5}) for
a model of solid fuel combustion \cite{FKRT, MY}. 
The type of the AHB in many conductance-base models
(including the original HH model) can be switched from sub- to supercritical by
varying parameters in physiologically relevant ranges \cite{RM}. Therefore, the results of this
paper are relevant to explaining firing patterns generated by this class of models.

\noindent {\bf Acknowledgments.}
One of the authors (GM) would like to thank Victor Roytburd for useful conversations.
This work was partially supported by the National Science Foundation 
under Grant No. 0417624.

\section*{Appendix. Parameter values. } 
\label{sec:A}
In this appendix, we list the expressions of various functions, which appear on the
right hand sides of (\ref{1}) and (\ref{2}):
\begin{eqnarray*}
\tau_s(v)={1\over \alpha_s(v)+\beta_s(v)},\; s\in\{h,n\}, &
c_\infty(v)={\alpha_c(v) \over \alpha_c(v)+\beta_c(v)},\; c\in\{m,n,h\}, \\
\alpha_m(v)={0.1(25-v)\over \exp\{ {25-v\over 10}\} -1}, & \beta_m(v)=4\exp\{{-v\over 18}\} \\
\alpha_n(v)={0.01(10-v)\over\exp\{{10-v\over 10}\}-1}, & \beta_n(v)=0.125\exp\{{-v\over 80}\}\\
\alpha_h(v)=0.07\exp\left\{{-(v-8)\over 20}\right\}, &  \beta_h(v)={1\over\exp\{ {38-v\over 10}\}+1}
\end{eqnarray*}
Here,  $v$, the membrane potential, is in mV. The gating variables $n$ and $h$ are nondimensional.
The values and the units of the parameter values in (\ref{1}) and (\ref{2}), which were used
to generate Fig. \ref{f.1}-\ref{f.5} are given in the following table:
\begin{center}
\begin{tabular}{|r|r||r|r||r|r||r|r|}
\hline
$I_{ap}$& $0.6\; \mu\mbox{A/cm}^2$ &
$\bar \tau_{h}$ & $1$ &
$\bar g_{Na}$ & $120\; \mbox{mS/cm}^2$ &
$\bar g_{K}$&$36\; \mbox{mS/cm}^2$ \\
$\bar g_{L}$&$0.3\; \mbox{mS/cm}^2$ &
$E_{Na}$&$115\; \mbox{mV}$ &
$E_{K}$&$-12\; \mbox{mV}$ &
$E_{L}$&$10.599\; \mbox{mV}$\\
\hline
\end{tabular}
\end{center}
The time constant $\bar\tau_n$ is viewed as a control parameter. We used $\bar\tau_n=42.925$
to generate Fig. \ref{f.1} and Fig. \ref{f.2}. 
For the values of  $\bar\tau_n$ used for other plots, we refer the reader to
the captions of the corresponding figures.

\end{document}